\documentclass[a4paper, 10pt, conference]{ieeeconf}      % Use this line for a4 paper

\IEEEoverridecommandlockouts                              % This command is only needed if 
\usepackage{graphics} % for pdf, bitmapped graphics files
\usepackage{epsfig} % for postscript graphics files
\usepackage{times} % assumes new font selection scheme installed
\usepackage{amsmath} % assumes amsmath package installed
\usepackage{amssymb}  % assumes amsmath package installed

\usepackage{mathrsfs}

\usepackage{pdflscape}
\usepackage{pgfplots}
 
\usetikzlibrary{arrows.meta}

\usepackage[hidelinks]{hyperref}

\DefineNamedColor{named}{green}{RGB}{0,180,0}

\DeclareFontFamily{U}{mathx}{\hyphenchar\font45}
\DeclareFontShape{U}{mathx}{m}{n}{<-> mathx10}{}
\DeclareSymbolFont{mathx}{U}{mathx}{m}{n}
\DeclareMathAccent{\widebar}{0}{mathx}{"73}

\usepackage{accents}

\usepackage{bm}

\usepackage{setspace}
\title{LSR-Net: Learning the Forward Evolution Operator for Nonlinear Fluid Dynamics}

\author{Qian Hou, Sutrisno, Yuqing Li and Zecheng Gan% <-this % stops a space
\thanks{$^{1}$Thrust of Advanced Materials and Guangzhou Municipal Key Laboratory of Materials Informatics, The Hong Kong University of Science and Technology (Guangzhou), Guangdong, China.
       {\tt\small qhou637@connect.hkust-gz.edu.cn}. Both Q. Hou and Z. Gan acknowledge the financial support from the Natural Science Foundation of China (Grant No. 12201146) and the Natural Science Foundation of Guangdong Province (Grant No. 2023A1515012197).}%
\thanks{$^{2}$Dept. of Mathematics, Universitas Diponegoro, Jalan Prof. Soedarto, Tembalang 50275, Semarang, Indonesia
       {\tt\small s.sutrisno @live.undip.ac.id}. This author thanks Universitas Diponegoro for the funding through RKU research grant contract no. 306-197/UN7.D2/PP/V/2026.}%
\thanks{$^{3}$ School of Mathematical Sciences, Key Laboratory of MEA \& Shanghai Key Laboratory of
PMMP, East China Normal University, Shanghai 200241, China.
       {\tt\small liyq@math.ecnu.edu.cn}.  Y. Li acknowledge the financial support from the Natural Science Foundation of China (Grant No. 12401667).}%
\thanks{$^{4}$ Thrust of Advanced Materials and Guangzhou Municipal Key Laboratory of Materials Informatics, The Hong Kong University of Science and Technology (Guangzhou), Guangdong, China;
and Department of Mathematics, The Hong Kong University of Science and Technology, Hong Kong
SAR, China.
       {\tt\small Corresponding author, zechenggan@hkust-gz.edu.cn}}%
}

\makeatletter
\def\@maketitle{
  \newpage
    \overrideIEEEmargins
  \vspace*{-19mm}   
  \begin{center}
  {\LARGE \bfseries \@title \par}
  \vskip 1em
  {\large \lineskip .5em
  \begin{tabular}[t]{c}
  \@author
  \end{tabular}\par}
  \end{center}
}
\makeatother
\begin{document}

\maketitle

%temporary add page number
% \thispagestyle{plain}
% \pagestyle{plain}

%%%%%%%%%%%%%%%%%%%%%%%%%%%%%%%%%%%%%%%%%%%%%%%%%%%%%%%%%%%%%%%%%%%%%%%%%%%%%%%%
\begin{abstract}
We introduce the Long-Short-Range Neural Network (LSR-Net), a novel neural operator architecture designed for data-driven forward evolution modeling, and extends it to the prediction of nonlinear fluid dynamics. 
LSR-Net learns the evolution operator of a dynamical system solely from pairs of initial and future state snapshots, which splits the learnable integral kernel into long-range (LR) and short-range (SR) components within stacked network blocks. 
While the SR component uses standard convolutions to capture local dynamics, the LR component employs a sum-of-exponentials (SOE) representation. This allows for the efficient computation of global interactions as a trainable Fourier multiplier, reducing computational complexity to $\mathcal O(n \log n)$ where $n$ is the number of pixels in an input snapshot and requiring only a few parameters per channel. 
LSR-Net is evaluated on three challenging 2D benchmarks: the coupled Burgers equation, the wave equation with a spatially varying coefficient, and the nonlinear shallow water equation {(SWE)}. 
Results demonstrate that LSR-Net significantly outperforms the baseline short-range network (SR-Net) as well as FNO and DeepONets in predictive accuracy, achieving substantially lower relative errors by effectively capturing both local fine-scale structures and crucial global pattern interactions.
\end{abstract}

\section{Introduction}
Data-driven forward evolution modeling aims to discover the underlying dynamical laws of physical systems directly from observational data, circumventing the need for first-principles derivation. In this context, operator learning has emerged as a powerful paradigm for learning the solution operators of partial differential equations (PDEs). Unlike traditional methods that approximate solutions for fixed parameters, operator learning frameworks learn mappings between infinite-dimensional function spaces, enabling them to predict the evolution of a system from an initial condition to a future state based solely on paired observation snapshots \cite{kovachki2023neural, Michalowska2024Neural, Jia2025Operator}.

Recent advances in operator learning have produced two prominent architectures: the Fourier Neural Operator (FNO)~\cite{Li2021} and DeepONet~\cite{Lu2021DeepONet}. 
FNO learns a neural operator by parameterizing the integral kernel directly in Fourier space, achieving resolution-invariance and computational efficiency through fast Fourier transforms. DeepONet, inspired by the universal approximation theorem for operators, consists of branch and trunk networks to learn operators from data. These methods have demonstrated remarkable success in learning complex PDE dynamics and provide flexible, model-agnostic frameworks for forward evolution modeling across various scientific and engineering applications.

Despite their successes, existing operator learning methods often struggle to capture long-range dependencies in dynamical systems, which can lead to inaccurate long-term predictions~\cite{li2020fourier, li2023long, nayak2025ti}. The challenge is particularly acute in pattern-forming systems where global interactions and local fine-scale structures coexist and co-evolve. To address these limitations, we propose the Long-Short-Range Neural Network (LSR-Net), which explicitly decomposes the learnable kernel into short-range and long-range components. By leveraging a sum-of-exponentials (SOE) representation for the long-range kernel, our method efficiently captures global interactions through a trainable Fourier multiplier with minimal parameter overhead, while standard convolutions handle local dynamics. This hybrid architecture enables LSR-Net to automatically discover hidden interaction patterns directly from data, achieving superior predictive accuracy across multiple challenging nonlinear fluid dynamics benchmarks.

\section{The Long-Short-Range Neural Network}

The goal of LSR-Net is to learn the time evolution of a dynamical system, where the underlying model remains unknown -- which can also be understood as a data-driven forward evolution modeling problem.
Given an initial system state $\phi_0=\phi(\mathbf{x},0)\in \mathcal{X}$, where $\mathcal{X}$ denotes the function space of system states, the objective is to predict the future state $\phi_t=\phi(\mathbf{x},t)$ after a finite time interval.

Assume that a set of paired observations $\{\phi_0^{(i)}, \phi_T^{(i)}\}_{i=1}^N$ is available at times $t=0$ and $t=T$.
The task is to approximate the evolution operator by a parameterized mapping
\[
G_\theta^T : \mathcal{X} \rightarrow \mathcal{X},
\]
where $\theta$ denotes the trainable parameters of the model.
The optimal parameters $\theta^\star$ are obtained by minimizing the mean squared error (MSE) between the predicted and reference states at time $T$:
\begin{equation}
\theta^\star
=
\arg\min_{\theta}
\frac{1}{N}
\sum_{i=1}^{N}
\left\|
G_\theta^T[\phi_0^{(i)}]
-
\phi_T^{(i)}
\right\|^2 .
\end{equation}
Once the evolution operator is learned, long-term predictions can be generated by repeatedly applying the operator to the current state, i.e.,
\[
\phi_0 \rightarrow \phi_T \rightarrow \phi_{2T} \rightarrow \cdots .
\]

Learning evolution operators is generally more challenging than standard function regression.
Physics-informed neural networks (PINNs)~\cite{Raissi2019} incorporate governing equations into the training process, but they are typically designed for fixed system parameters.
Operator learning approaches such as the Fourier Neural Operator (FNO)~\cite{Li2021} and DeepONet~\cite{Lu2021DeepONet} provide more flexible frameworks for learning mappings between function spaces.
However, these methods may still struggle to capture long-range dependencies, which can lead to inaccurate long-term predictions~\cite{li2023long, nayak2025ti}.

\begin{figure}[htbp]
\centering
\includegraphics[width=0.48\textwidth]{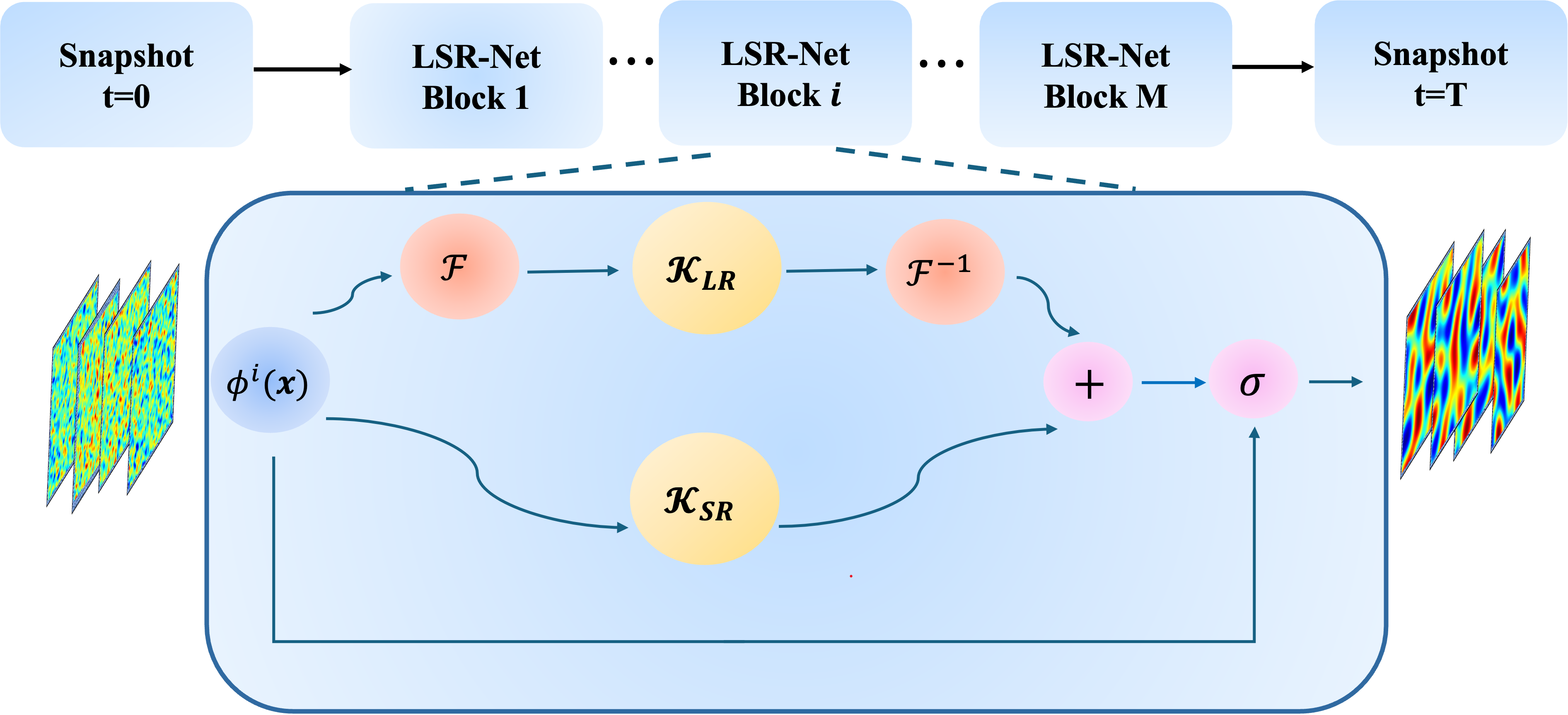}
\caption{
Architecture of LSR-Net. 
Starting from the input snapshot at $t=0$, the network applies $M$ stacked LSR-Net blocks to predict the system state at time $t=T$. 
Each block combines long-range (LR) and short-range (SR) learnable integral kernels followed by nonlinear activation functions. 
The LR convolution is efficiently computed in Fourier space using the proposed sum-of-exponentials (SOE) kernel representation. 
A residual connection is introduced in each block to improve training stability.
}
\label{fig:arch_NN}
\end{figure}

The LSR-Net architecture is illustrated in Fig.~\ref{fig:arch_NN}, which consists of $M$ stacked LSR-Net blocks.
Each block combines two types of learnable integral kernels: a long-range (LR) kernel that captures global interactions and a short-range (SR) kernel that models local dynamics.
The outputs are then passed through a nonlinear activation function $\sigma$.
To improve training stability, a residual connection is introduced in each block.
This design alleviates gradient exploding or vanishing during backpropagation.

Starting from the input state at $t=0$, the network applies $M$ sequential LSR-Net blocks to produce the predicted state at time $T$.
Denoting the transformation in each block by $\{ g_{\theta_i} \}_{i=1}^M$, the learned evolution operator can be written as
\begin{equation}
G^T_{\bm{\theta}}[\phi] 
=
(\sigma \circ g_{\theta_M}) 
\circ 
(\sigma \circ g_{\theta_{M-1}})
\circ 
\cdots 
\circ 
(\sigma \circ g_{\theta_1})[\phi].
\end{equation}

Each operator $g_{\theta_i}$ is defined through an integral kernel transformation:
\begin{equation}
g_{\theta_i}[\phi^i(\mathbf{x})]
=
\int_{\Omega}
\mathcal{K}(\mathbf{x}-\mathbf{y};\theta_i)
\,\phi^i(\mathbf{y})\,d\mathbf{y}.
\label{kernel}
\end{equation}
Here $\mathcal{K}(\mathbf{x}-\mathbf{y};\theta_i)$ is a learnable nonlocal kernel parameterized by $\theta_i$, and $\phi^i(\mathbf{x})$ denotes the intermediate output of the $(i-1)$-th block.
Consequently, $\bm{\theta}=\{\theta_i\}_{i=1}^M$ represents the full set of trainable parameters.
In LSR-Net, the activation function $\sigma$ is chosen as \texttt{Tanh}, which has been shown to be effective in modeling particle and fluid interfaces~\cite{tanaka2000simulation}.
When physical models are available, tailored activation functions may also be derived analytically under certain conditions~\cite{Lan2023DOSnet}.

To capture both non-local interactions and local fine-scale structures in pattern formation, we decompose the kernel $\mathcal{K}$ into long-range (LR) and short-range (SR) components:
\begin{equation}
\mathcal{K} = \mathcal{K}_{\text{LR}} + \mathcal{K}_{\text{SR}} .
\end{equation}
Following common practices in convolutional neural networks (CNNs), the SR component is modeled using a standard convolution layer with a small kernel size (typically {$5\sim7$}). 
In contrast, capturing LR interactions using conventional convolutions is more challenging, as it usually requires either very large kernels or deep network stacks.

To address this limitation, we introduce a sum-of-exponentials (SOE) representation for the LR kernel:
\begin{equation}\label{eq:KLR}
\mathcal K_{LR}(\mathbf{x} - \mathbf{y}; \mathbf\theta_{LR})
\approx
\sum_{l=1}^{P}
\beta_l e^{-\alpha_l |\mathbf{x}-\mathbf{y}|},
\end{equation}
where $\mathbf\theta_{LR}=\{\alpha_l,\beta_l\}_{l=1}^P$ are learnable parameters and $P$ denotes the number of LR convolution channels.
The SOE representation is motivated by approximation theory~\cite{devore2009multiscale}, which states that smooth kernels can be approximated to arbitrary accuracy by weighted sums of exponential functions. 
Equation~\eqref{eq:KLR} implicitly assumes translational invariance of the evolution operator. 
If this symmetry does not hold, the SOE representation can be extended to higher-dimensional forms~\cite{Jiang2015}.
Unlike many previous studies that determine the optimal parameters for a predefined kernel form~\cite{jiang2008efficient,gao2022kernel,lin2025weighted}, our framework does not assume any prior knowledge of the kernel structure. 
Instead, all parameters are learned directly from data, enabling the model to automatically discover hidden long-range interaction patterns.

The SOE representation also enables efficient computation of LR convolutions. 
Using the Fourier transform pair
\[
\left\{e^{-\alpha|x|},~\frac{2\alpha}{\alpha^2+k^2}\right\},
\]
the LR convolution can be evaluated in Fourier space via the convolution theorem:
\begin{equation}\label{LR_kernel}
\int_{\Omega}
\mathcal{K}_{LR}(\mathbf{x}-\mathbf{y};\mathbf\theta_{LR})
\phi(\mathbf{y})\,d\mathbf{y}
=
\mathcal{F}^{-1}
\left[
\widehat{\mathcal{K}}_{LR}(\mathbf{k})
\cdot
\widehat{\phi}(\mathbf{k})
\right],
\end{equation}
where $\mathcal{F}^{-1}$ denotes the inverse Fourier transform. 
In Fourier space, the LR kernel becomes a trainable Fourier multiplier
\begin{equation}
\widehat{\mathcal{K}}_{LR}(\mathbf{k})
=
\sum_{i=1}^{P}
\beta_i
\left(
\frac{\alpha_i}{k^2+\alpha_i^2}
\right).
\end{equation}
In many applications, LR interactions may exhibit oscillatory behavior. 
This can be incorporated by extending the exponents into the complex domain~\cite{gao2022kernel,lin2025weighted}. 
Using the Fourier transform pair
\[
\left\{e^{-\alpha|x|}\cos(sx),~
\frac{\alpha}{\alpha^2+(k+s)^2}
+
\frac{\alpha}{\alpha^2+(k-s)^2}
\right\},
\]
the Fourier multiplier can be written as
\begin{equation*}
\widehat{\mathcal{K}}_{LR}^{os}(\mathbf{k})
=
\sum_{i=1}^{P}
\beta_i
\left[
\frac{\alpha_i}{(k+s_i)^2+\alpha_i^2}
+
\frac{\alpha_i}{(k-s_i)^2+\alpha_i^2}
\right].
\end{equation*}
When $s_i=0$, this expression reduces to the non-oscillatory case. 
Even for oscillatory kernels, only three trainable parameters are required per LR channel.

Although the two sides of Eq.~\eqref{LR_kernel} are mathematically equivalent, their computational costs differ substantially. 
Direct evaluation of the LR convolution requires $\mathcal{O}(n^2)$ operations, where $n$ is the number of pixels in an input snapshot. 
By performing the computation in Fourier space, the cost is reduced to $\mathcal{O}(n\log n)$ through three steps: (1) compute the fast Fourier transform (FFT) of the input field, (2) apply point-wise multiplication with the Fourier multiplier, and (3) transform back using the inverse FFT.

\section{Numerical results}
\subsection{Data Generation}
To evaluate the performance of our model, we generate synthetic datasets by numerically solving several nonlinear fluid dynamics models. Specifically, following~\cite{rosofsky2023pino}, we consider three benchmark systems:
\begin{itemize}
    \item 2D Burgers equation (vectorized form with coupled fields $u(x,y,t)$ and $v(x,y,t)$)
    \item 2D wave equation with spatially varying coefficient
    \item 2D shallow water system (used here as a {SWE}-type nonlinear wave benchmark)
\end{itemize}
Each entry corresponds to a trajectory and contains a tensor of solution fields over time:

\[
\mathbf{U} \in \mathbb{R}^{T \times N_x \times N_y \times C},
\]
where
\begin{itemize}
\item $T$ is the number of saved time steps,
\item $N_x=N_y=128$ is the spatial resolution,
\item $C$ is the number of physical channels.
\end{itemize}
For Burgers and wave equations $C=1$, while for the shallow water system $C=3$ corresponding to $(h,u,v)$.
All simulations are performed on a uniform periodic grid of size $128 \times 128$. Initial conditions are sampled from Gaussian random fields (GRFs) with Matérn covariance.

\subsection{Model Setting}

We summarize the model and dataset configurations used for the three PDE benchmarks studied in this work: 2D Burgers equation, 2D wave equation with spatially varying coefficients, and 2D nonlinear shallow water system ({SWE}-type).

% After training, the model accuracy is evaluated using the relative mean squared error (RMSE) defined as:
{After training, the model accuracy is evaluated using the relative $L^2$ error, defined as:}
\begin{equation}
    {\mathrm{Relative}\ L^2\ \mathrm{Error}} = \frac{ \sum_{i=1}^{N_{test}} ({\hat \phi}^{(i)} - {\phi}^{(i)})^2}{ \sum_{i=1}^{N_{test}} {\phi^{(i)}}^2},
\end{equation}
where ${\hat \phi}^{(i)}$ is LSR-Net prediction while ${\phi}^{(i)}$ is ground truth from full simulations, and $N_{test}$ is number of test samples.

\begin{table}[h!]
\centering
\caption{Summary of model and dataset settings for all PDE benchmarks.}
\resizebox{\columnwidth}{!}{
\begin{tabular}{lccc}
\hline
\textbf{Parameter} & \textbf{Burgers 2D} & \textbf{Wave 2D} & \textbf{{SWE 2D}} \\
\hline
Dataset size & 500 samples & 500 samples & 500 samples \\
Domain & $[0,1)^2$ & $[0,1)^2$ & $[0,1)^2$ \\
Num. layers & 3 & 3 & 3 \\
Feature channels & 7 & 7 & 9 \\
Kernel size & 7 & 11 & 11 \\
Batch size & 1 & 4 & 4 \\
Learning rate & 0.001 & 0.001 & 0.001 \\
Optimizer & \begin{tabular}[t]{@{}l@{}}Adam\\ $\epsilon=1\mathrm{e}{-8}$\end{tabular} 
          & \begin{tabular}[t]{@{}l@{}}AdamW~\cite{loshchilov2017decoupled}\\ $\epsilon=1\mathrm{e}{-6}$\end{tabular} 
          & \begin{tabular}[t]{@{}l@{}}Adam\\ $\epsilon=1\mathrm{e}{-8}$\end{tabular} \\
Scheduler & \begin{tabular}[t]{@{}l@{}}step\_size=16\\ $\gamma=0.5$\end{tabular} 
          & \begin{tabular}[t]{@{}l@{}}step\_size=16\\ $\gamma=0.5$\end{tabular} 
          & \begin{tabular}[t]{@{}l@{}}step\_size=8\\ $\gamma=0.1$\end{tabular} \\
Epochs & 70 & 70 & 70 \\
\hline
\end{tabular}
}
\label{tab:model_setting}
\end{table}

\noindent
\textbf{Notes:} 
\begin{itemize}
    \item The Burgers 2D system is a coupled vector field $(u,v)$ with periodic boundary conditions and initial conditions sampled from Gaussian random fields.
    \item The Wave 2D system uses a spatially varying wave speed $c(x,y)$ as an additional input field to the PINO, with independent GRF initial displacement $u_0(x,y)$.
    \item The {SWE} 2D system models the nonlinear shallow water equations with total height $\eta(x,y,t)$, initial perturbation $\eta_0$, and zero initial velocity fields $u_0=v_0=0$.
    \item All models are implemented in PyTorch. Training and inference are performed on a single device MPS.
\end{itemize}

\subsection{2D Coupled Burgers Equation}

We consider the two-dimensional vectorized Burgers equation, which consists of two coupled velocity fields $u(x,y,t)$ and $v(x,y,t)$:

\begin{equation}
\begin{cases}
\partial_t u + u \, \partial_x u + v \, \partial_y u = \nu (\partial_{xx} u + \partial_{yy} u), \\
\partial_t v + u \, \partial_x v + v \, \partial_y v = \nu (\partial_{xx} v + \partial_{yy} v),
\end{cases}
\end{equation}
where $(x,y) \in [0,1)^2$, $t \in [0,T]$, and $\nu= 0.01$ is the viscosity coefficient. This system is periodic in both spatial directions and exhibits nonlinear coupling between $u$ and $v$ through the advection terms.
Initial conditions for both velocity components are independently sampled from a Gaussian random field (GRF) with correlation length $l = 0.1$ and standard deviation $\sigma = 0.2$, i.e.,
\[
u(x,y,0),\, v(x,y,0) \sim \mathcal{GRF}(l,\sigma).
\]
The equations are solved using a spectral solver with time step $\Delta t = 1\times 10^{-4}$ up to the final time $T = N_{\text{steps}} \cdot \Delta t$, and snapshots are recorded every $\Delta t_{\text{save}} = 0.01$ time units. The resulting dataset contains $N_{\text{samples}} = 500$ independent realizations of the coupled dynamics, each stored in a tensor of shape $[2, t_{\text{steps}}, N_x, N_y]$, where the first dimension indexes the velocity components $[u,v]$.

\begin{figure}[t]
\centering
\includegraphics[width=0.95\linewidth]{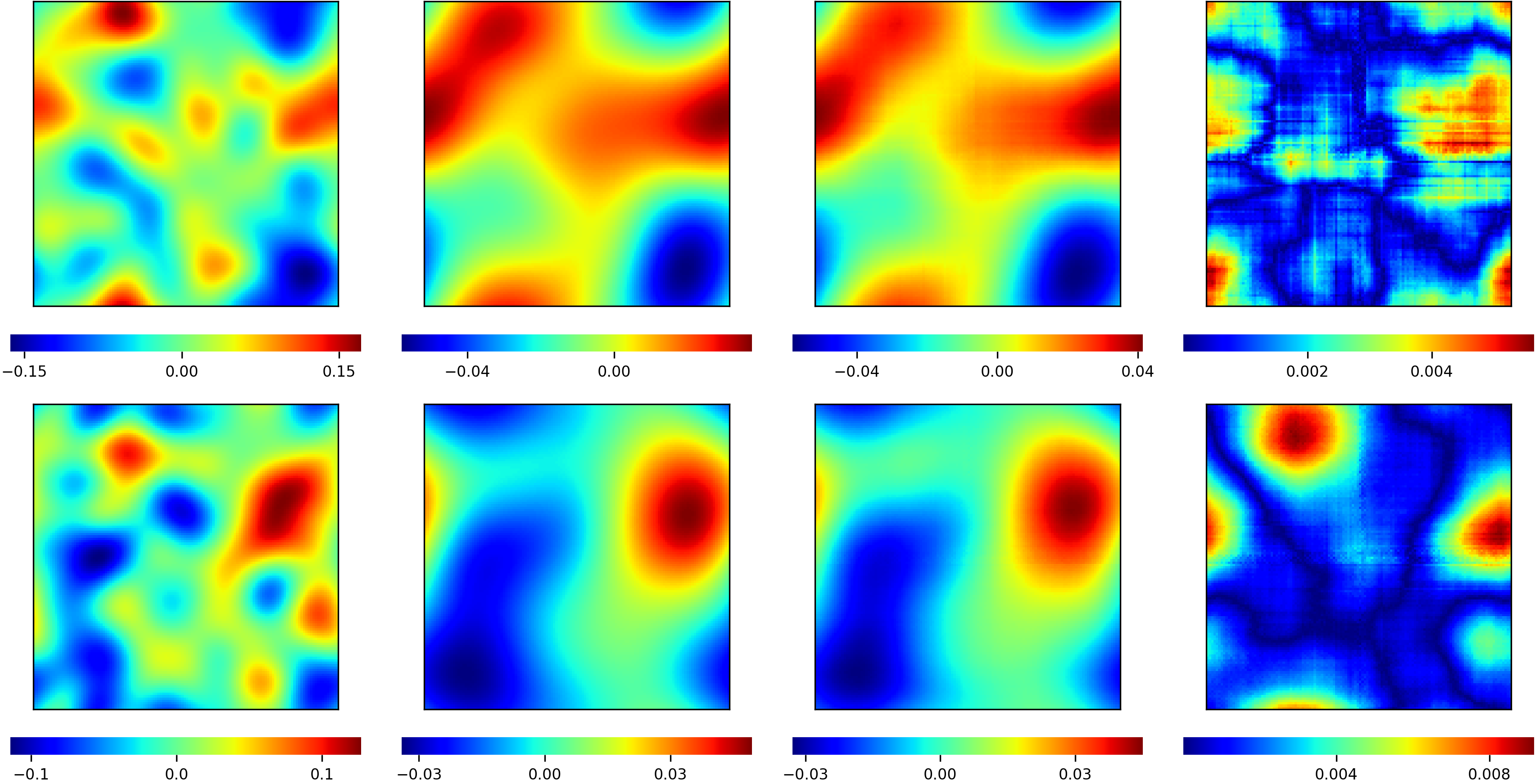}
\includegraphics[width=0.95\linewidth]{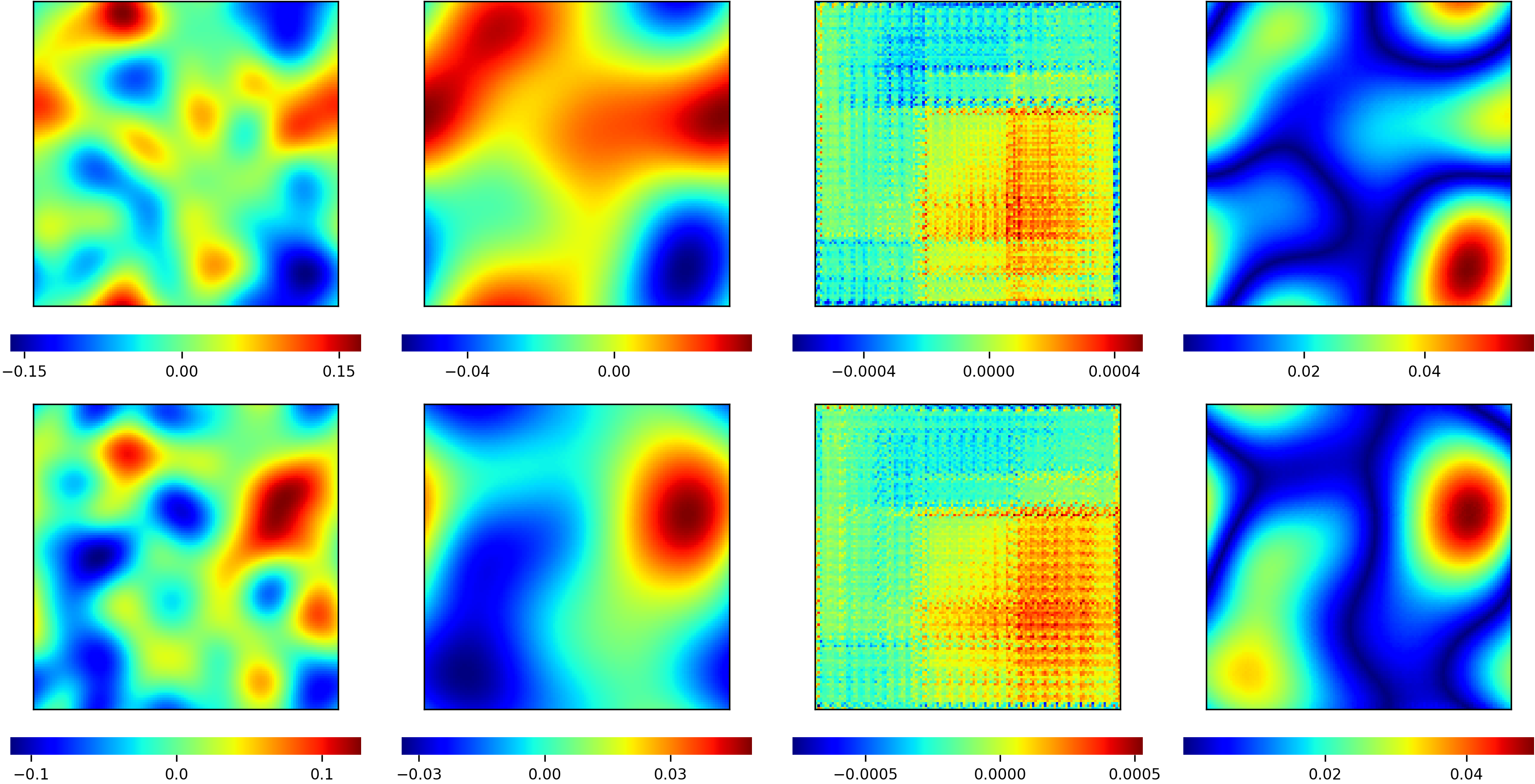}
\caption{
Comparison of prediction results on the 2D Burgers equation. 
Each row visualizes one velocity component. The four columns show the initial condition, the ground truth at $t=T$, the model prediction, and the absolute prediction error. 
The first two rows correspond to the results produced by LSR-Net, while the last two rows correspond to SR-Net. 
}
\label{fig:burger_compare}
\end{figure}

 We compare the proposed LSR-Net with a baseline SR-Net, which omits the long-range SOE kernel and only retains the short-range convolution. Since the 2D Burgers equation involves two velocity components, $u$ and $v$, we visualize the prediction results for both variables. In Fig.~\ref{fig:burger_compare}, the four columns correspond to the initial condition, the ground truth at $t=T$, the network prediction, and the corresponding absolute error.
The first two rows show the results produced by LSR-Net, while the last two rows correspond to SR-Net. It is clear that LSR-Net exhibits significant advantages for both coupled variables compared to SR-Net. LSR-Net demonstrates significant advantages for both coupled variables compared to SR-Net. The SR-Net not only struggles to capture the underlying patterns but also produces ambiguous predictions, likely due to its relatively small kernel size of 7. This highlights the accuracy gains provided by the long-range SOE kernel, which enables LSR-Net to capture global flow structures more effectively than networks with smaller kernels..This comparison demonstrates that incorporating the long-range SOE kernel allows LSR-Net to better capture global flow structures, resulting in more accurate predictions and lower errors than SR-Net.

\subsection{2D Wave Equation with Spatially Varying Coefficient}

In this experiment, we consider a two-dimensional wave equation with a spatially varying wave speed:

\begin{equation}
\begin{aligned}
\partial_{tt} u(x,y,t) &= c(x,y)^2 \, (\partial_{xx} u + \partial_{yy} u), \\
u(x,y,0) &= u_0(x,y), \\
\partial_t u(x,y,0) &= 0, \\
(x,y) &\in [0,1)^2, \quad t \in [0,T],
\end{aligned}
\end{equation}
where $c(x,y)$ denotes the non-constant wave speed and  is sampled from a Gaussian random field (GRF) with correlation length $l_c = 0.5$ and standard deviation $\sigma = 1$.  The initial displacement field $u_0(x,y)$ is independently sampled from another GRF with correlation length $l = 0.1$ and the same standard deviation. The equation is solved on a periodic domain using a spectral solver with time step $\Delta t = 10^{-4}$ up to final time {$T=5$}, and snapshots are recorded at fixed intervals. 

\begin{figure}[t]
\centering
\includegraphics[width=1\linewidth]{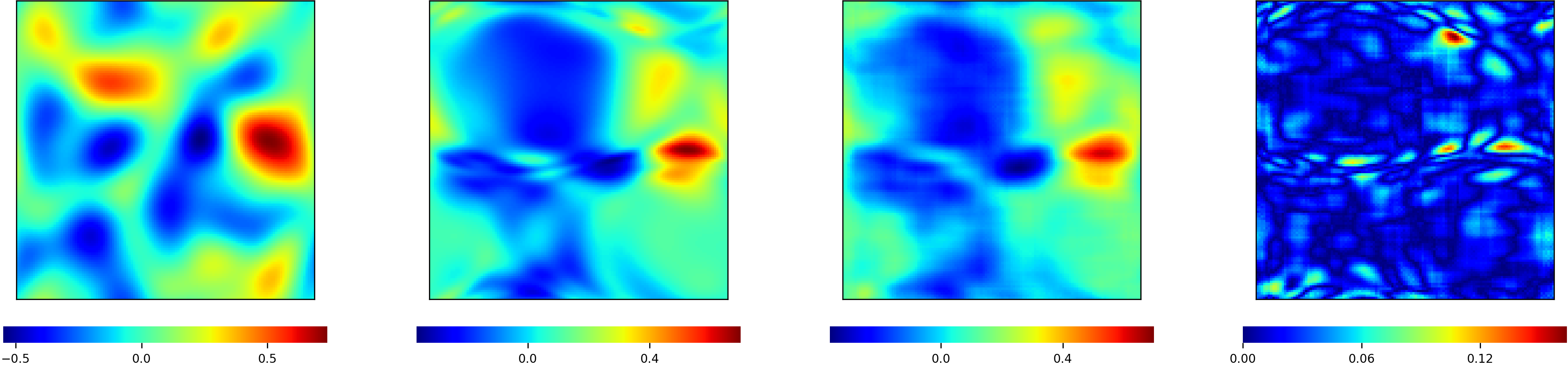}
\vspace{0.3em}
\includegraphics[width=1\linewidth]{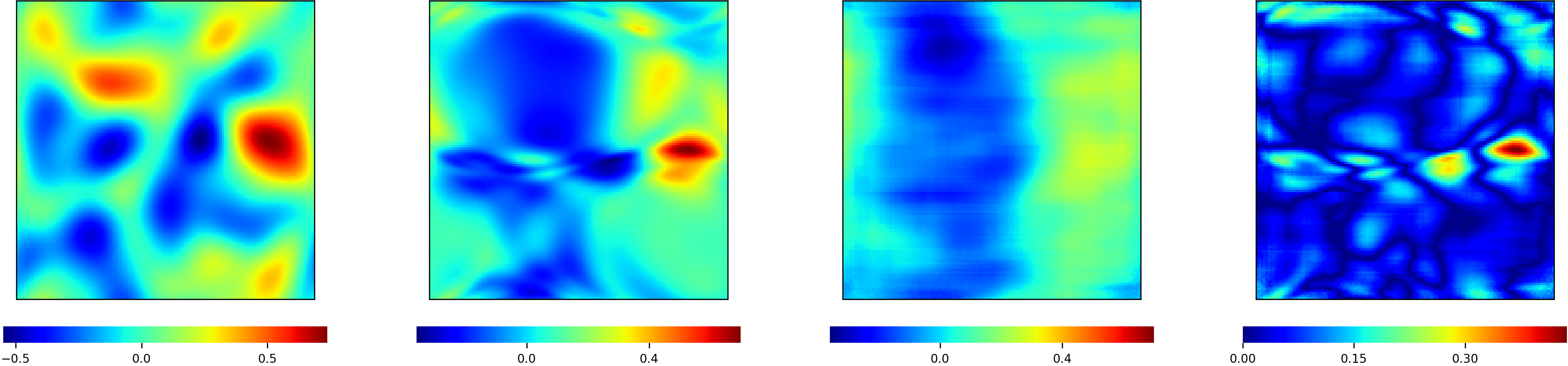}
\caption{
Prediction comparison for the wave equation with spatially varying coefficient. 
The wave field $u$ is a scalar variable, while the wave speed $c(x,y)$ varies spatially, which makes the prediction task more challenging. 
Each row corresponds to one neural operator model: the first row shows the results of LSR-Net and the second row shows the results of SR-Net. 
The four columns represent the initial condition, the ground truth at $t=T$, the predicted solution, and the absolute prediction error. 
}
\label{fig:wave_compare}
\end{figure}

Results are shown in Fig.~\ref{fig:wave_compare}. 
Compared to the Burgers equation, this problem is more challenging, since the wave speed $c(x,y)$ \emph{varies spatially}, leading to heterogeneous propagation dynamics. 
Such spatial variability introduces complex wave interactions, making accurate prediction significantly more difficult. 
Each row in Fig.~\ref{fig:wave_compare} visualizes the solution of the scalar field $u$. 
The first row corresponds to the prediction produced by LSR-Net, while the second row corresponds to SR-Net. 
The four columns represent the initial condition, the ground truth at $t=T$, the model prediction, and the corresponding prediction error, respectively. 
The visualization highlights that LSR-Net produces \emph{more accurate predictions} and better captures \emph{sharp wave structures}, whereas SR-Net exhibits larger errors in regions with strong spatial variations. 
Overall, the results clearly demonstrate the advantage of LSR-Net in handling spatially varying wave speeds and preserving the sharpness of complex patterns.

\subsection{Nonlinear Shallow Water Equations (2D)}

We examine the network performance on the nonlinear shallow water equations. 
The total fluid column height is $\eta(x,y,t)$, given by a mean value of 1 plus some initial perturbation. 
The initial horizontal velocity fields $u(x,y,t)$ and $v(x,y,t)$ are set to zero.

The governing equations are
\begin{equation*}
\begin{aligned}
\partial_t \eta + \partial_x (\eta u) + \partial_y (\eta v) &= 0, \\
\partial_t (\eta u) + \partial_x \Big(\eta u^2 + \frac{g}{2} \eta^2 \Big) + \partial_y (\eta u v) &= \nu (\partial_{xx} u + \partial_{yy} u), \\
\partial_t (\eta v) + \partial_x (\eta u v) + \partial_y \Big(\eta v^2 + \frac{g}{2} \eta^2 \Big) &= \nu (\partial_{xx} v + \partial_{yy} v),
\end{aligned}
\end{equation*}
with initial conditions $\eta(x,y,0) = \eta_0(x,y)$, $u(x,y,0) = 0$, $v(x,y,0) = 0$, 
and periodic boundary conditions $(x,y) \in [0,1)^2$, for $t \in [0,1]$. 
Here, $g = 1$ is the gravitational acceleration and $\nu = 0.002$ is the viscosity coefficient to prevent shocks. 
The initial height field $\eta_0(x,y)$ is sampled from a Gaussian random field with correlation length $l = 0.1$ and variance $\sigma = 0.2$. 
Simulations are performed using time step $\Delta t = 10^{-3}$ up to final time $T=1$.

% \begin{figure}[t]
% \centering
% \includegraphics[width=0.9\linewidth]{pic/LSR_SWE_test_0_3_diff.png}
% \vspace{0.3em}
% \includegraphics[width=0.9\linewidth]{pic/SR_SWE_test_0_3_diff.png}
% \caption{
% Prediction comparison for the nonlinear shallow water system. 
% The system contains three variables: the height field $h$ and the velocity components $u$ and $v$. 
% Therefore, the visualization consists of six rows. The first three rows show the predictions from LSR-Net for $h$, $u$, and $v$, respectively, while the last three rows correspond to SR-Net. 
% The four columns represent the initial condition, the ground truth at $t=T$, the predicted solution, and the absolute prediction error. 
% Although the initial velocity fields are zero, LSR-Net successfully captures the induced velocity dynamics through the coupling with the height field. In contrast, SR-Net produces much larger errors for the velocity components, indicating its difficulty in learning the coupled nonlinear dynamics.
% }
% \label{fig:swe_compare}
% \end{figure}

\begin{figure*}[t]  % 使用 figure* 跨双栏
\centering
\begin{minipage}[b]{0.48\linewidth}
  \centering
  \includegraphics[width=\linewidth]{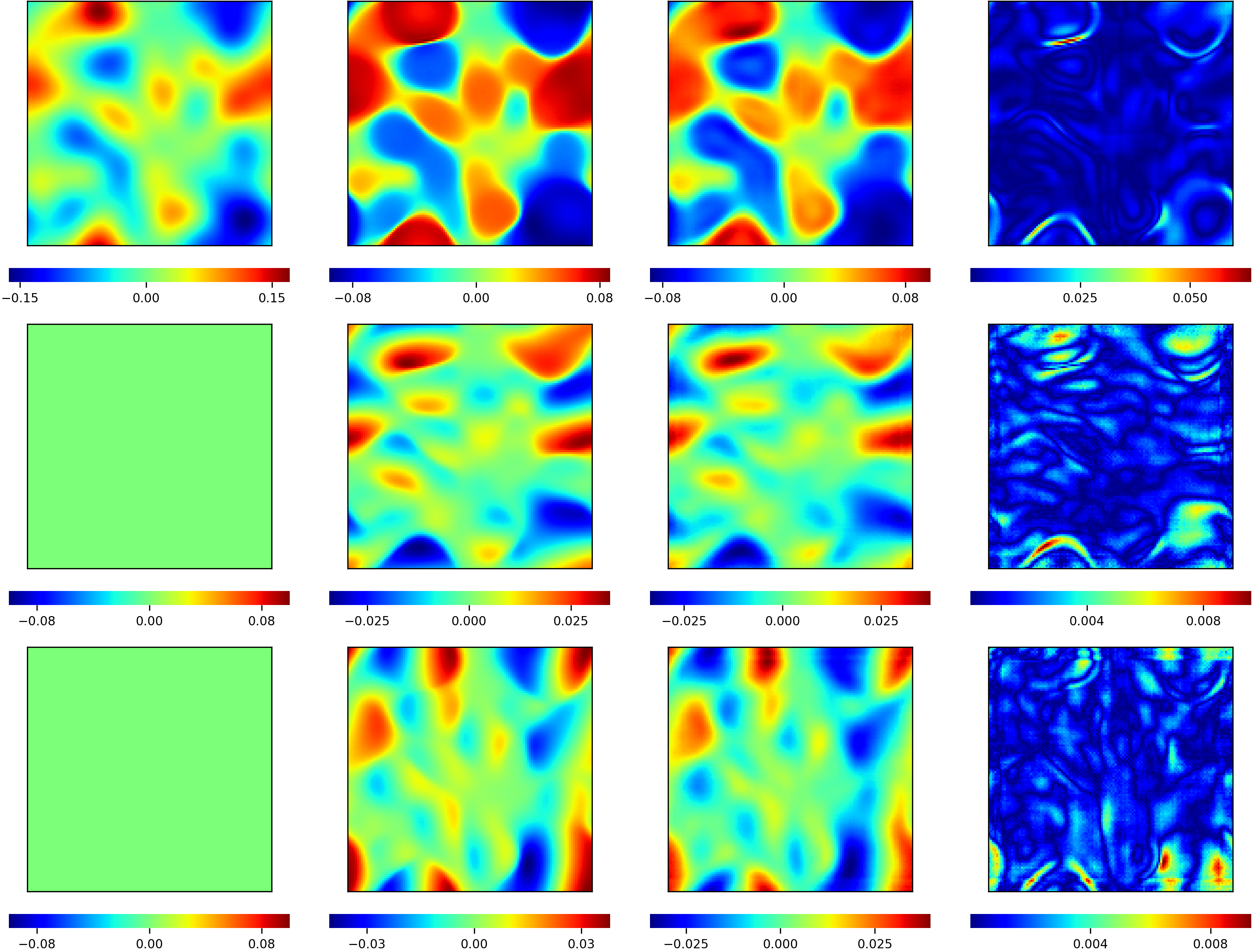}
  {{SWE} results from LSR-Net}
\end{minipage}
\hfill
\begin{minipage}[b]{0.48\linewidth}
  \centering
  \includegraphics[width=\linewidth]{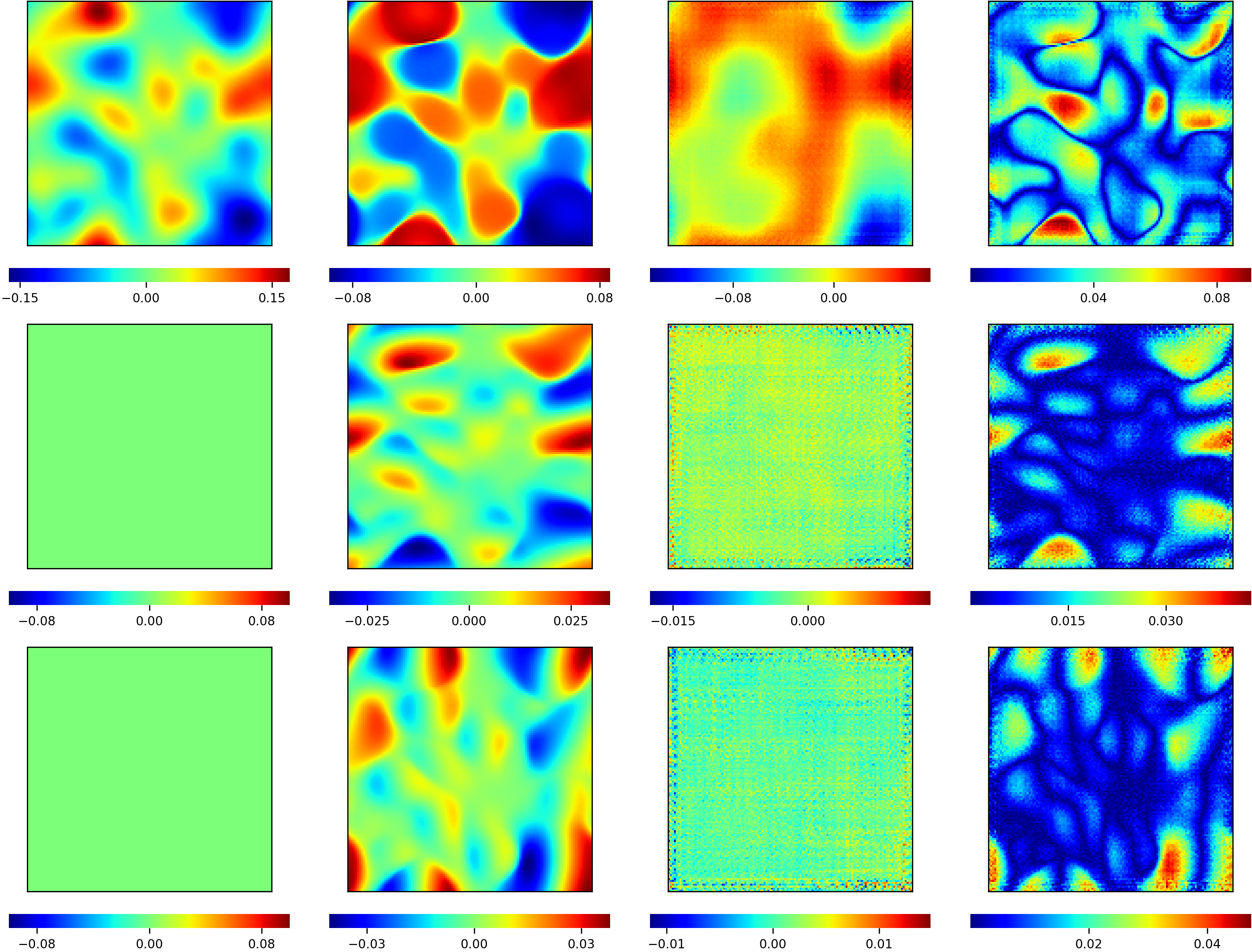}
  {{SWE} results from SR-Net}
\end{minipage}

\caption{
Prediction comparison for the nonlinear shallow water system. 
The system contains three variables: the height field $h$ and the velocity components $u$ and $v$. 
The four columns in each subfigure represent the initial condition, the ground truth at $t=T$, the predicted solution, and the absolute prediction error. 
}
\label{fig:swe_compare}
\end{figure*}

Results are shown in Fig.~\ref{fig:swe_compare}. 
The nonlinear shallow water system involves three coupled physical variables: the height field $\eta$ and the two velocity components $u$ and $v$. 
This multi-variable coupling introduces \emph{three main difficulties} for prediction: 
1) the system is multi-dimensional,  
2) the initial velocity components $u$ and $v$ are \emph{zero}, making it very challenging to learn their evolution, and  
3) the variables are strongly coupled through nonlinear dynamics.  

The visualization contains six rows: the first three rows correspond to LSR-Net predictions for $\eta$, $u$, and $v$, respectively, while the last three rows correspond to SR-Net. 
The four columns represent the initial condition, the ground truth at $t=T$, the model prediction, and the corresponding prediction error. 
Despite the simple initial velocities, LSR-Net successfully captures the induced velocity dynamics through the coupling with $\eta$ and produces accurate predictions for all three variables. 
In contrast, SR-Net fails to accurately predict all three variables, including the height field $\eta$, as well as the velocity components $u$ and $v$, resulting in significantly larger errors across the system.
These results demonstrate that LSR-Net effectively handles the multi-variable coupling and the challenge of zero initial velocities.

\begin{table}[h!]
\centering
\caption{Comparison of the {relative $L^2$ errors} achieved by LSR-Net and SR-Net on three PDE benchmarks.}
\begin{tabular}{lcc}
\hline
\textbf{PDE} & \textbf{SR-Net} & \textbf{LSR-Net} \\
\hline
2D Coupled Burgers & 0.999926 & 0.01388 \\
2D Wave Equation   & 0.429731 & 0.03060 \\
{SWE System}      & 0.475511 & {0.01807} \\
\hline
\end{tabular}
\label{tab:rmse_comparison}
\end{table}

Table~\ref{tab:rmse_comparison} reports the relative errors (RMSE) of SR-Net and LSR-Net on the three PDE benchmarks. 
It is immediately clear that LSR-Net achieves substantially lower errors across all cases. 
For the 2D coupled Burgers equation, SR-Net nearly fails with an RMSE of 0.9999, while LSR-Net reduces it to 0.0139, 
demonstrating its ability to accurately capture the coupled dynamics. 
In the 2D wave equation with spatially varying wave speed, SR-Net produces a relatively large RMSE of 0.4297 
due to the heterogeneous propagation patterns, whereas LSR-Net achieves a much lower error of 0.0306, 
effectively capturing sharp wave structures. 
Similarly, for the {SWE system}, SR-Net has an RMSE of 0.4755, while LSR-Net attains an extremely low error of 0.0181. 
Overall, these results clearly illustrate the superiority of LSR-Net in learning long-range interactions 
and accurately predicting complex PDE dynamics compared with a short-range network.

\subsection{{Multi-step inference comparison with different methods}}

To evaluate the long-term prediction capability of the proposed framework, we compare LSR-Net with three representative baseline methods, including SR-Net, FNO, and DeepONet, on three PDE benchmarks: the 2D coupled Burgers equation, the 2D wave equation, and the SWE system.

All methods are evaluated under an autoregressive rollout setting up to $5T$. For the 2D coupled Burgers equation and the SWE system, the rollout interval is set to $T=0.2$, resulting in long-term inference up to $5T=1.0$. For the 2D wave equation, the rollout interval is set to $T=1$, and autoregressive inference is performed up to $5T=5$.
At each rollout step, the prediction from the previous step is recursively fed back into the model as the input for the next-step prediction. The prediction accuracy is quantified using the relative $L^2$ error.
For model setting, LSR-Net and SR-Net are same in Table \ref{tab:model_setting}. For FNO, the key parameters are hidden layer and number of frequency modes are  4 and and for 2D coupled Burgers, 7 and 4 for SWE system, 10 and 4for  2D wave equation. For DeepONet, the latent basis dimension is set to $p=10$, and the number of trunk network layers is set to $4$ for all three PDE systems.

Tables~\ref{tab:burgers_rollout}, \ref{tab:swe_rollout}, and \ref{tab:wave_rollout} present the autoregressive multi-step rollout relative $L^2$ errors for LSR-Net, SR-Net, FNO, and DeepONet on the 2D coupled Burgers equation, the SWE system, and the 2D wave equation, respectively.

For the 2D coupled Burgers equation, LSR-Net consistently achieves the lowest prediction errors across all rollout horizons, with significantly slower error accumulation compared with FNO. In contrast, SR-Net rapidly collapses during autoregressive inference, while DeepONet exhibits substantially larger prediction errors throughout the rollout process.

A similar trend is observed for the SWE system. LSR-Net maintains the best overall prediction accuracy during long-term rollout, while FNO shows moderate error growth as the inference horizon increases. DeepONet and SR-Net both suffer from unstable autoregressive behavior and fail to preserve accurate long-term dynamics.

For the 2D wave equation, all methods experience noticeable error accumulation during long-term rollout. LSR-Net achieves the best short-term prediction accuracy at early rollout stages ($\leq 3T$), demonstrating strong local approximation capability. However, it is clear that due to the varying coefficients of the wave equation, the operator is no longer isotropic, thus as the rollout horizon increases, accumulated errors lead to degraded long-term prediction performance.

Overall, the proposed LSR-Net demonstrates competitive autoregressive prediction capability across multiple PDE systems and consistently outperforms the ablated SR-Net variant, highlighting the importance of the LR component for stable long-term operator learning.

\begin{table}[t]
\centering
\caption{Comparison of multi-step rollout relative $L^2$ errors on the 2D coupled Burgers equation.}
\small
\label{tab:burgers_rollout}
\resizebox{\columnwidth}{!}{
\begin{tabular}{|c|c|c|c|c|c|}
\hline
Method & $T$ & $2T$ & $3T$ & $4T$ & $5T$ \\
\hline

LSR-Net 
& \textbf{0.00061} 
& \textbf{0.00132} 
& \textbf{0.00253} 
& \textbf{0.00425} 
& \textbf{0.00652} \\
\hline

SR-Net 
& 1.00021 
& 1.00029 
& 1.00037 
& 1.00047 
& 1.00058 \\
\hline

FNO 
& 0.00312 
& 0.00659 
& 0.01115 
& 0.01722 
& 0.02514 \\
\hline

DeepONet 
& 0.86706 
& 0.88453 
& 0.92775 
& 0.96634 
& 0.99764 \\
\hline

\end{tabular}
}
\end{table}

\begin{table}[t]
\centering
\caption{Comparison of multi-step rollout relative $L^2$ errors on the SWE system.}
\small
\label{tab:swe_rollout}
\resizebox{\columnwidth}{!}{
\begin{tabular}{|c|c|c|c|c|c|}
\hline
Method & $T$ & $2T$ & $3T$ & $4T$ & $5T$ \\
\hline

LSR-Net 
& \textbf{0.01395}
& \textbf{0.04017}
& \textbf{0.07067}
& \textbf{0.10346}
& \textbf{0.13792} \\
\hline

SR-Net 
& 1.00079
& 1.00080
& 1.00084
& 1.00088
& 1.00093 \\
\hline

FNO 
& 0.01547
& 0.04724
& 0.08789
& 0.13593
& 0.19103 \\
\hline

DeepONet 
& 0.96283
& 1.00873
& 1.04133
& 1.03780
& 1.03277 \\
\hline

\end{tabular}
}
\vspace{-0.5cm}
\end{table}

\begin{table}[t]
\centering
\caption{Comparison of multi-step rollout relative $L^2$ errors on the 2D wave equation.}
\small
\label{tab:wave_rollout}
\resizebox{\columnwidth}{!}{
\begin{tabular}{|c|c|c|c|c|c|}
\hline
Method & $T$ & $2T$ & $3T$ & $4T$ & $5T$ \\
\hline

LSR-Net 
& \textbf{0.00092}
& \textbf{0.03980}
& \textbf{0.24167}
& 0.67059
& 1.91164 \\
\hline

SR-Net 
& 0.00417
& 0.05145
& 0.27185
& 0.68912
& 1.17940 \\
\hline

FNO 
& 0.00300
& 0.05938
& 0.27559
& \textbf{0.62816}
& 0.97727 \\
\hline

DeepONet 
& 0.60926
& 0.58023
& 0.58751
& 0.65133
& \textbf{0.74370} \\
\hline

\end{tabular}
}
\vspace{-0.5cm}
\end{table}

\section{Conclusion}

In this work, we propose the long-short-range neural network (\emph{LSR-Net}), tailored for prediciton in complex pattern formation scenarios. This model-free approach requires only two sets of early-stage snapshots. The key novelty of LSR-Net lies in the integration of a long-range convolution module, which efficiently captures global interactions through a trainable Fourier multiplier derived from the sum-of-exponentials (SOE) ansatz in kernel approximation theory. Notably, this requires only 3–4 trainable parameters per long-range convolution channel. By combining the long-range SOE kernel with a standard short-range convolutional kernel, LSR-Net is able to effectively explore and learn the evolution of complex patterns. We evaluate LSR-Net on three challenging PDE benchmarks: the 2D coupled Burgers equation, the 2D wave equation with spatially varying wave speed, and the {SWE system}. Compared with a short-range network that lacks the SOE long-range kernel, the results demonstrate that LSR-Net consistently achieves superior predictive accuracy across all cases, particularly for long-term predictions.
In the future, we plan to extend LSR-Net to broader nonlinear dynamic systems, such as anisotropic systems or irregular geometries, as well as operators act on general manifolds.

\bibliographystyle{IEEEtran}
\bibliography{reference}  

\end{document}